\pdfoutput=1

\documentclass[11pt]{article}

\usepackage[]{acl}

\usepackage{times}
\usepackage{latexsym}

\usepackage[T1]{fontenc}

\usepackage[utf8]{inputenc}

\usepackage{microtype}

\usepackage{inconsolata}

\usepackage{csquotes}
\usepackage{amsmath}
\usepackage{amsthm}
\usepackage{hyperref}
\usepackage{bm}
\usepackage{bbm}
\usepackage{mathtools}
\usepackage{booktabs}
\usepackage[ruled, vlined, linesnumbered]{algorithm2e}
\usepackage{subcaption}
\usepackage[switch]{lineno}
\usepackage{xcolor} 
\usepackage{multirow}
\usepackage{mathrsfs} 
\usepackage{balance}
\usepackage{multicol}
\usepackage{makecell}
\usepackage{amssymb}

\usepackage{floatrow}
\newfloatcommand{capbtabbox}{table}[][\FBwidth]

\newcommand{\huimin}[1]{\textcolor{black}{#1}}
\newcommand{\ours}{{GPT-FedRec}\xspace}

\title{Federated Recommendation via Hybrid Retrieval Augmented Generation}

\author{Huimin Zeng \quad Zhenrui Yue \quad Qian Jiang \quad Dong Wang \\
Unversity of Illinois at Urbana-Champaign \\
\{\texttt{huiminz3, zhenrui3, qianj3, dwang24\}@illinois.edu}}

\begin{document}
\maketitle
\begin{abstract}
Federated Recommendation (FR) emerges as a novel paradigm that enables privacy-preserving recommendations. However, traditional FR systems usually represent users/items with discrete identities (IDs), suffering from performance degradation due to the data sparsity and heterogeneity in FR. On the other hand, Large Language Models (LLMs) as recommenders have proven effective across various recommendation scenarios. Yet, LLM-based recommenders encounter challenges such as low inference efficiency and potential hallucination, compromising their performance in real-world scenarios. To this end, we propose \textbf{\ours}, a federated recommendation framework leveraging ChatGPT and a novel hybrid \huimin{Retrieval Augmented Generation} (RAG) mechanism. \ours is a two-stage solution. The first stage is a hybrid retrieval process, mining ID-based user patterns and text-based item features. Next, the retrieved results are converted into text prompts and fed into GPT for re-ranking. Our proposed hybrid retrieval mechanism and LLM-based re-rank aims to extract generalized features from data and exploit pretrained knowledge within LLM, overcoming data sparsity and heterogeneity in FR. In addition, the RAG approach also prevents LLM hallucination, improving the recommendation performance for real-world users. Experimental results on diverse benchmark datasets demonstrate the superior performance of \ours against state-of-the-art baseline methods.
\end{abstract}

\section{Introduction}
\label{sec:intro}
Recommendation systems play a vital role in aiding users to discover pertinent content of their interests, such as online commerce \cite{ying2018graph}, social media \cite{fan2019graph}. The prevailing strategy in developing recommendation systems involves extracting users' personalized preferences from their historical data. However, the concerns on data privacy have prompted strict regulations on data governance (e.g., GDPR \footnote{General Data Protection Regulation:https://gdpr-info.eu/}). Such regulations limit the development of modern recommendation systems.

To address this issue, federated recommendation (FR) emerges as a new paradigm facilitating privacy-preserving recommendations \cite{ammad2019federated,zhang2024transfr}. In FR, a central server stores a global recommendation model, while a set of clients contain local private data. These clients collaboratively train the global model without sharing their private data. Note that the concept "client" does not necessarily refer to a single human user, but can also represent a local data server with a small dataset. \textbf{For clarity, we consistently use "client" to represent a federated client, and use "user" to denote a human user of a recommender system.} 

\begin{figure}[t]
\centering
\includegraphics[trim=6.4cm 3.5cm 6.4cm 2.4cm, clip, width=0.9\textwidth]{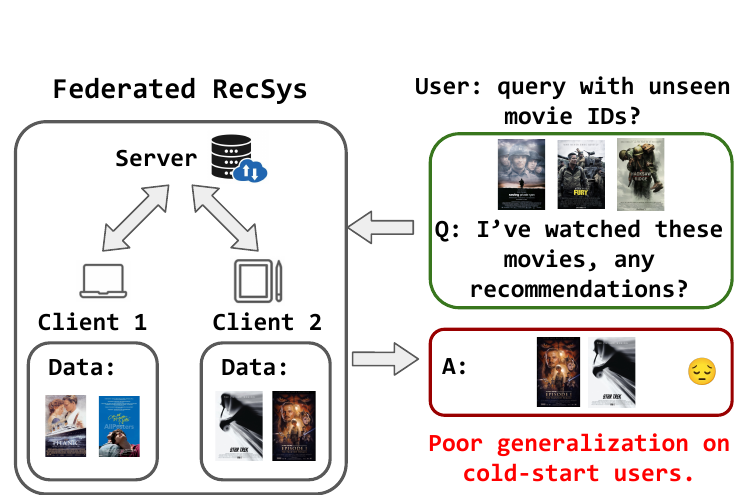}
\caption{Under data heterogeneity and data sparsity, traditional ID-based recommenders fail to recommend correct items to the cold-start users.}
\label{fig:example}
\vspace{-0.2cm}
\end{figure}

However, traditional FR systems usually represent users/items with discrete identities \cite{zhang2024transfr}. As such, they may suffer from degraded performance, due to the data sparsity and heterogeneity in the federated setting \cite{wu2022fedcl,zhang2024transfr}. In FR, the clients might only contain the data of a few or even a single user (data sparsity). Training an ID-based recommender on such sparse data is prone to overfitting \cite{zhang2024transfr}. Moreover, data sparsity may also trigger data heterogeneity. That is, the item scopes of local datasets are only subsets of the entire item scope, and different local clients may have different item scopes (non-i.i.d. data). As shown in Figure~\ref{fig:example}, a test user's data contains novel items never seen in the local clients (i.e., a cold-start user). Consequently, the FR system could not make meaningful recommendations. Since the locally trained models may never see certain items during training, aggregating such sparsely-trained models results in poor generalization \cite{zhang2024transfr}.

\begin{figure*}[t]
\centering
\centering
\includegraphics[trim=0.6cm 4.2cm 1.2cm 4.3cm, clip, width=0.9\textwidth]{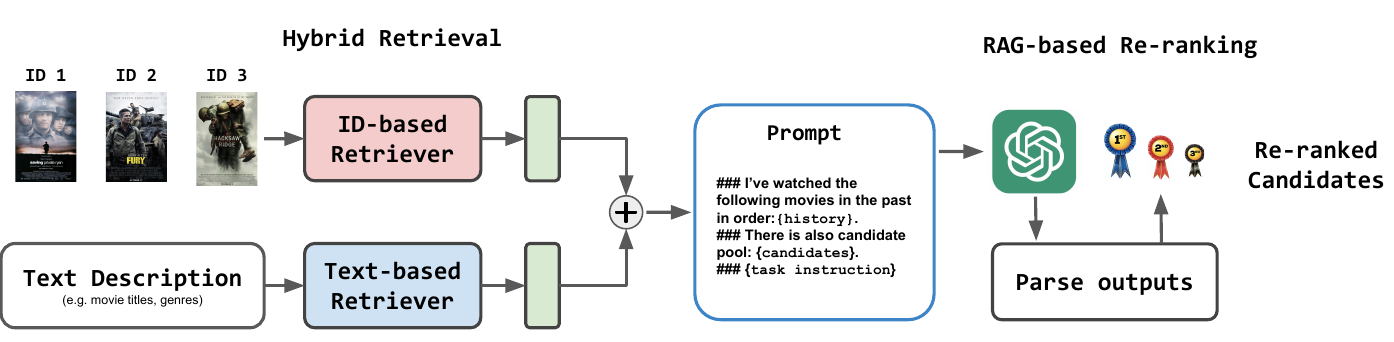}
\caption{\ours. The first stage involves a hybrid retrieval with an ID-based retriever and a text-based retriever. The second stage performs re-ranking via the retrieved results and RAG of the prompted GPT.}
\label{fig:overview}
\end{figure*}


On the other hand, large language models (LLMs) exhibit strong generalization abilities in diverse tasks, thanks to the extensive knowledge learned from the massive-scale real-world data. Recently, LLMs as recommenders also demonstrate impressive performance in different recommendation tasks \cite{harte2023leveraging, li2023bookgpt, bao2023tallrec, zhang2023recommendation, yue2023llamarec}. \huimin{Therefore, employing LLMs for FR has the potential to address data sparsity and data heterogeneity.} Yet, existing LLM-based recommenders suffer from issues like incomplete recommendations, low inference efficiency and potential hallucination \cite{yue2023llamarec,li2023large}, impairing their applicability in real-world scenarios. Besides, in FR, it is neither affordable nor feasible for local clients to finetune LLMs with limited computational resources and data. In addition, the next-word-generating nature of LLMs might generate irrelevant words in terms of recommendation, leading to incomplete recommendations and low inference efficiency \cite{yue2023llamarec}. \huimin{Finally, as a consequence of hallucination, LLM-based recommenders may recommend non-existing or make-up items, compromising user experience in real-world applications. }

To this end, in this work, \huimin{we propose \ours, a novel and effective two-stage framework for federated recommendation. Leveraging both the generalized features within training data and pretrained knowledge within LLMs (e.g., ChatGPT), \ours not only provides a complete solution in a data sparse and heterogeneous FR setting, but also outperforms existing methods with improved recommendation performance.} 

As shown in Figure~\ref{fig:overview}, \ours operates in two stages. The first stage employs a novel hybrid retrieval mechanism to generate recommendation candidates using generalized data features. We propose to adopt small-scale ID-based retrievers to capture the ID-based user patterns, and dense text-based retrievers to extract robust and generalized semantic features from item descriptions. The rationale behind this design is: \huimin{while the ID-based user history is informative in terms of representing user-item dynamics, the textual item descriptions (e.g., titles, categories) contain more generalized semantic features. Leveraging such generalized features improves generalization of the recommendation, especially in data-sparse and heterogeneous FR settings \cite{zhang2024transfr}}. Consider the movie recommendation in Figure~\ref{fig:example}. The descriptions of these movies provide more information for the recommender: the movies watched by the new user could be depicted in texts like "war, action movies", which is semantically closer to the "sci-fi, action movies" on client 2, compared to the "love, romantic movies" on client 1. In the second stage, \huimin{the retrieved candidates are fed into an LLM for re-ranking. This process leverages the pretrained generalized knowledge of LLM, thereby further improving the generalization of the recommendation. Moreover, conditioning LLM re-rank on the first-stage results effectively prevents hallucination. This enhances recommendation performance for real-world users.} \huimin{Finally, due to space limit, we narrow down the scope of this work to the sequential recommendation, for its superior performance over traditional recommendater systems (e.g., matrix factorization).} We summarize the contributions of our paper as follows\footnote{We adopt publicly available datasets and release the code at \url{https://github.com/huiminzeng/GPT-FedRec.git}.}:
\begin{enumerate}

\item To the best of our knowledge, \ours is the first FR framework that uses hybrid RAG and LLMs. More importantly, \ours withstands the critical challenge of data sparsity and data heterogeneity in FR, achieving promising performance.

\item Technically, \ours employs arbitrary ID-based retriever and arbitrary text-based retriever to perform the hybrid retrieval. Moreover, as a RAG method, \ours does not require the finefuning of the LLMs.

\item We evaluate \ours in the FR setting. On different benchmark datasets, experimental results suggest that \ours \huimin{achieves considerable improvements in recommendation performance}, while outperforming state-of-the-art baselines.
\end{enumerate}

\section{Related Work}
\label{sec:rela}

\paragraph{Federated Recommendation.} 
Existing federated recommendation (FR) systems are usually ID-based models: users/items are represented with unique identities (IDs). For instance, FCF \cite{ammad2019federated}, FedRec \cite{lin2020fedrec}, FedMF \cite{chai2020secure} extend classic ID-based matrix factorization into federated mode. In comparison, \citet{wu2022fedcl} train ID-based sequential recommenders to domain-disentangled features in a federated fashion, achieving better performance. \citet{luo2022towards} aggregates ID-based model with a client utility-aware protocol for better federated recommendation. However, in heterogeneous settings, these ID-based FR methods could not generalize well to cold-start users, because the items in the cold-start user data might never be seen during training \cite{zhang2024transfr}. Targeting at this issue, \citet{zhang2024transfr} proposed a text-based FR solution, namely TransFR. Trained using textual features, TransFR demonstrates better generalization on heterogeneous FR settings. However, TransFR suffers from data sparsity and is not capable of handling long user-item sequences. Compared to existing FR methods, the hybrid retrieval mechanism in \ours is designed to address the data sparsity and heterogeneity in FR. Unlike TransFR, \ours is capable of modeling long sequences. 

\paragraph{Natural Language for Recommendation.}
The inherent generality of textual features empowered the generalizability and transferability of recommenders. For instance, to achieve better transferability, \citet{hou2023learning,hou2022towards} train sequential recommenders using language-model-encoded item texts. Similary, in \cite{li2023text,geng2022recommendation}, pretrained language models are finetuned using item descriptions and then used as recommenders. These text-based recommender systems demonstrate state-of-the-art generalizability in different recommendation scenarios. However, they usually require rich textual data to enable finefuning and could not handle long user histories. On the other hand, LLMs as recommenders have gained increased attention for LLMs' impressive generalization ability. The mainstream LLM-based recommenders exploits the pretrained knowledge within LLMs to perform next-item recommendation \cite{hou2023large,sileo2022zero,sun2023chatgpt,he2023large}.  For example, \citet{he2023large} employees LLMs as conversational agents to understand user preferences and improve recommendation. Another stream of LLM-based recommendation designs tuning strategies tailored for specific subtasks (e.g., rating prediction) to improve recommendation performance \cite{chen2023palr,kang2023llms,yue2023llamarec}. However, existing LLM-based recommenders mainly generate recommendation candidates in an autoregressive fashion. As such, the generated content might be irrelevant to the items of interests or even are hallucinated, leading to undesired performance in real-world applications. As such, we design a retrieval augmented recommendation framework, where the RAG approach effectively reduces hallucination and improves recommendation performance.

\section{Preliminaries}
\label{sec:preliminary} 

\paragraph{Data.} 
\huimin{We adopt the sequential recommendation setting to illustrate the data format.} In sequential recommendation, a data point is the historical data of a user: a sequence of interacted items ${x}$ (sorted by timestamps) within her history. A sequence ${x}$ is a list of items $[x_1, x_2, ..., x_l]$ of length $l$. Each element in ${x}$ belongs to the item scope $\mathcal{I}$ that contains all items: $x_i \in \mathcal{I}$. The goal of a recommender is to predict the next user-item interaction $x_{l+1} \in \mathcal{I}$ based on the user history ${x}$. In our experiments, $x_{l+1}$ is used as ground truth $y$ (i.e., $y = x_{l+1}$).

\paragraph{Model.} 
An ID-based recommender directly takes item sequences as input, and maps them into new items as recommendations. Given an input sequence ${x}$, an ID-based recommender computes a score vector over the item scope $\mathcal{I}$, and recommends items with highest probabilities. In comparison, a text-based recommender usually transforms discrete item IDs into descriptions: transforming a sequence of items ($x$) into a sequence of descriptions. After trained on such text sequences, the text-based recommender generates textual IDs (i.e., IDs in text format) or item titles as the final recommendation.

\paragraph{Federated Recommendation.}
In FR, a central server stores a global recommender, while a set of local clients store their respective datasets. The goal of FR is to collaboratively train the global recommender without clients sharing their private data. Assume there are $K$ clients in an FR application. Each client contains a local dataset $\mathcal{D}^{k}$ with $|\mathcal{D}^{k}|$ item sequences. In FR, the item scopes covered by local datasets may be smaller than the entire item scope: $\mathcal{I}^{k} \subseteq \mathcal{I}$ and $\cup_{k} \mathcal{I}^{k} \subseteq \mathcal{I}$ (data sparsity). In addition, some local clients may contain unique items only present in their datasets (data heterogeneity). The test user can also introduce new, unseen items (i.e., cold-start users). \huimin{As in Figure~\ref{fig:example}, the clients contain disjoint movies, and the test user data involves new, unseen movies.} Therefore, the challenges of data sparsity, data heterogeneity and non-generalizabilty on cold-start users drive the development of \ours. \textbf{For simplicity, unless specified, notations without client index $k$ represent the data of an arbitrary client.} 

\section{Algorithm}
\label{sec:alg}
The overview of \ours is in Figure~\ref{fig:overview}. To address the aforementioned challenges, we firstly develop a hybrid retrieval mechanism using a small-scale ID-based retriever and a dense text-based retriever to retrieve generalized candidates (Section~\ref{sec:hybried_retrieval}). Then, based on the retrieved results, we establish a retrieval augmented recommendation pipeline using LLMs (Section~\ref{sec:llm}). 

\subsection{Hybrid Retrieval}
\label{sec:hybried_retrieval}
\paragraph{ID-based Retriever.} Given existing ID-based recommenders, any of them could be used by \ours to retrieve potential candidates. In our implementation, we choose LRURec \cite{yue2023linear} as the ID-based retriever for its state-of-the-art performance and light-weight design, avoiding extra communication costs in FR settings. 

Formally, we define the ID-based retriever as $f_{I}$, parameterized by $\theta_{I}$. As shown in Figure~\ref{fig:overview}, for a piece of user data, $f_{I}$ takes its interacted item sequence $x$ as input, i.e., $f_{I}(x)$. $f_{I}$ returns a vector of similarity scores over the item scope $\mathcal{I}$.  To train LRURec on each local client, we optimize the cross-entropy loss over the local dataset:
\begin{equation}
\label{eq:lru_ce} 
    \mathcal{L}_{ce} = \mathbb{E}_{({x}, y) \sim \mathcal{D}^{k}} [\mathcal{L}({f}_{I}({x}), y)].
\end{equation}

During federated training, the weights of $f_{I}$ are sent to the global server for aggregation. However, it is foreseeable that in a data sparse and data heterogeneous setting, the predicted scores returned by a local $f_{I}$ are skewed towards the items present in its local dataset. That is, a locally trained $f_{I}$ is prone to retrieve the items present its local datasets, rendering its unsatisfactory performance on test data with unseen new items.

\begin{figure*}[!bht]
\centering
\includegraphics[trim=0.6cm 4.2cm 1.2cm 4.3cm, clip, width=0.9\textwidth]{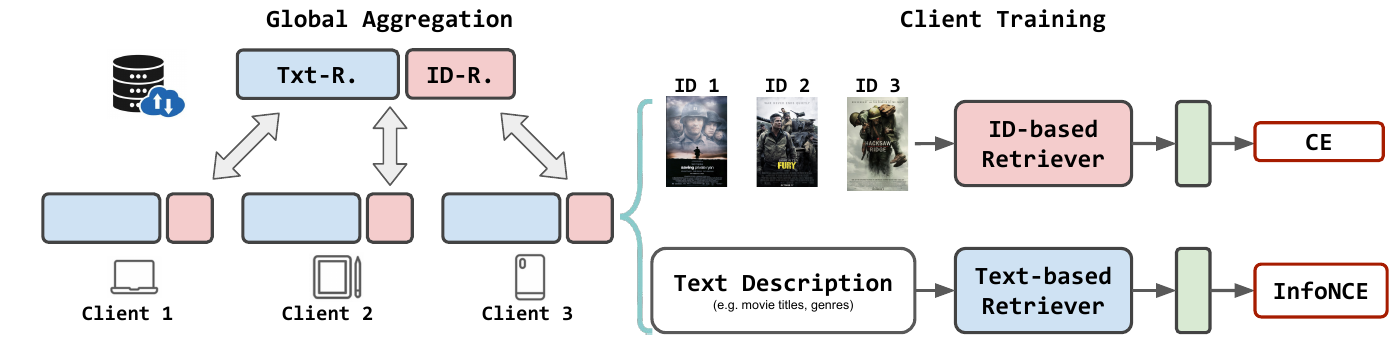}
\caption{The training and aggregation process of \ours. On clients, the ID-based retriever and the text-based retriever are trained using local data.}
\label{fig:train}
\end{figure*}

\paragraph{Text-based Retriever.} In the data sparse and data heterogeneous FR setting, the inadequacy of ID-based retriever motivates us to employ an additional dense, more generalizable text-based retriever to build \ours. To illustrate the necessity of using text-based retriever, consider the example in Figure~\ref{fig:example}. In this example, client 1, client 2 and the test user have mutually exclusive item scopes. This is a typical data heterogeneous case. In the discrete ID space, a movie with a specific ID can only appear in one party. An ID-based retriever would fail to capture the correlation between some movie IDs cross parties, leading to degraded performance. In a sharp contrast, in the text space, the movies across these parties may still share common, generalized features (e.g., "war, action movies" of the new user and "sci-fi, war movies" on client 2), despite the different movie IDs. As such, text-based retriever could capture such generalized features, overcoming data heterogeneity that defects ID-based retrievers. 

To this end, we propose to use E5 \cite{wang2022text} as the text-based retriever. E5 is a transformer-based language model, pretrained for text-retrieval tasks. Formally, we define the text-based retriever as $f_{T}$, parameterized by $\theta_{T}$. 

To adapt E5 into our FR application, we finetune it using text descriptions of items and the InfoNCE loss \cite{oord2018representation} on each local client. In particular, for each training sample $x=[x_1, x_2, ..., x_l]$ and its ground-truth item $y$, we firstly transform them into input texts $t$ and ground-truth text $t_y$ using the item meta data (e.g., titles, categories, genres of the items) using the E5 template.
\begin{equation}
\label{eq:e5_template}
\begin{split}
    & t = [t_1, t_2, ..., t_l], \quad \mathrm{where} \quad t_i = \texttt{Template}_{1}(x_i), \\
    & t_y = \texttt{Template}_{2}(y). 
\end{split}
\end{equation}
In Equation~\ref{eq:e5_template}, $\texttt{Template}_{1}$ is the E5 template with "query" as prefix, $\texttt{Template}_{2}$ is the E5 template with "passage" as prefix. \huimin{(More details about E5 templates in Appendix~\ref{app:1}.)} Then, with prepared text templates, $f_{T}$ is finetuned with the InfoICE loss:
\begin{equation}
\label{eq:infonce}
\begin{split}
    & \mathcal{L}_{info} = \\
    & -\frac{1}{|\mathcal{D}|}\sum_{m}^{|\mathcal{D}|}\mathrm{log}\frac{e^{s(q^{(m)}, p^{(m)})}}{e^{s(q^{(m)}, p^{(m)})} + \sum_{n}e^{s(q^{(m)}, \bar{p}^{(n)})}},  \\
    & \mathrm{where} \\
    & q = f_{T}(t), \;  p = f_{T}(t_y), \; \bar{p} = f_{T}{(\bar{y})}, \; \bar{y}\in \mathcal{I} \setminus y
\end{split}
\end{equation}
In Equation~\ref{eq:infonce}, we use superscript $(m)$ to index different users (not client). As for $s(q, p)$, it refers to a scoring function between the encoded query $q$ and an encoded positive passage $p$ or negative passage $\bar{p}$. In our implementation, the score function is cosine similarity scaled by a temperature $\tau$:
\begin{equation}
    s(p,q) = \mathrm{cos}(p, q) / \tau.
\end{equation}
By finetuning E5 using Equation~\ref{eq:infonce}, the text-based retriever learns to assign high similarity between user sequences and their ground-truth items in the text space. More importantly, since E5 is pretrained, it has already encoded prior knowledge of language patterns. Therefore, the text-based retriever is capable of extracting generalized textual features from item descriptions, despite the data sparsity and data heterogeneity across clients.

Finally, after training the ID-based retriever ($f^{k}_{I}$) and the text-based retriever ($f^{k}_{T}$) on each local client, we adopt FedAvg \cite{mcmahan2017communication} to perform global model aggregation, obtaining the global ID-based retriever and text-based retriever:
\begin{equation}
\label{eq:fed_avg}
\begin{split}
    & \theta^*_{(\cdot)} = \frac{1}{\sum_{k} |\mathcal{D}^{k}|}\sum_{k=1}^K |\mathcal{D}^{k}| \cdot 
    {\theta}^{k}_{(\cdot)} \\
    & \mathrm{s.t.} \quad {\theta}^{k}_{I} = \arg \min \mathcal{L}_{ce} (\mathcal{D}^{k}, f_{I}), \\
    & \qquad \; {\theta}^{k}_{T} = \arg \min \mathcal{L}_{info} (\mathcal{D}^{k}, f_{T}), \\
    & \qquad \; k \in \{1,...,K\}.
\end{split}
\end{equation}

\paragraph{Hybrid Retrieval.} To generate the retrieval results for any user sequence $x$, we compute a weighted sum of the normalized prediction scores returned by the aggregated ID-based retriever and text-based retriever through
the Tikhonov principle \cite{tikhonov1963solution}:
\begin{equation}
\label{eq:hybrid}
\begin{split}
    & \hat{P}_{hybrid} = \lambda \cdot \sigma(f_I(x)) + (1- \lambda) \cdot \sigma(f_{T}(t)),
\end{split}
\end{equation}
where $\sigma$ is the softmax opernation. Then, the items of top-N scores within $\hat{P}_{hybrid}$ are retrieved as candidates. These candidates form a candidate set $\hat{\mathcal{I}}$. $\hat{\mathcal{I}}$ is the result of the first stage of \ours.

We highlight that the candidate set $\hat{\mathcal{I}}$ is retrieved based on the hybrid scores $\hat{P}_{hybrid}$. Compared to traditional ID-based FR models, the hybrid scores incorporate both domain-generalized features from the text-based retriever and the representative ID-based user patterns. As such, \ours is equipped with generalization ability by design, and can overcome the data sparsity and data heterogeneity issue in FR applications, achieving better recommendation performance.

\subsection{Hybrid Retrieval Augmented Generation}
\label{sec:llm}
\paragraph{RAG-based Recommendation.} After the hybrid retrieval stage, \ours further employs an LLM to perform re-ranking among the retrieved candidates within $\hat{\mathcal{I}}$. This design is to exploit the pretrained knowledge encoded in LLMs, so that the generalization of recommendations is further enhanced. Moreover, the re-ranking process of the LLM is conditioned on the retrieved results, hallucination could be effectively avoided: \huimin{the LLM is explicitly instructed to only re-rank the candidates within $\hat{\mathcal{I}}$, and they are retrieved from real-world data during the first stage.} Finally, \huimin{for efficiency, we intend to prompt an LLM for recommendation without finetuning it}, and treat the LLM-based re-ranking process as retrieval augmented generation \cite{de2020autoregressive,tay2022transformer}. 

\paragraph{LLM.} In this work, we adopt GPT-3.5-Turbo \cite{ChatGPT} from OpenAI to build \ours. GPT-3.5-Turbo is a closed-source LLM with abilities of solving many complex tasks \cite{ChatGPT}. Since in a data sparse FR setting, it is infeasible to finetune LLM using such limited data, we select GPT-3.5-Turbo for its powerful zero-shot generalization ability.

To construct a text prompt for GPT-3.5-Turbo, inspired by \cite{hou2023large}, we start it with a description of both user history and candidate item using their titles and categories/genres. Then, an instruction to the task is appended to the description. In addition to the text prompts, we also feed system prompts to GPT, so that its mindset is adjusted to the concrete recommendation application (e.g., recommending movies or cosmetics). A simplifed prompt template looks like:

\begin{displayquote}
\#\#\# system: You are a helpful \{\texttt{role}\},  \{\texttt{role description}\}.

\#\#\# user: I've browsed the following items in the past in order: \{\texttt{history}\}. There is also candidate pool: \{\texttt{candidates}\}. \{\texttt{task instruction}\}. 
\end{displayquote}

In the above template, \texttt{role} and \texttt{role description} are replaced with a concrete real-world job (e.g., shopping assistant) and its job description (e.g., recommending products for customers). \texttt{history} is instantiated as item texts containing item titles, possibly with item categories/genres. As for \texttt{candidates}, it is replaced with the hybrid retrieval results. \texttt{task instruction} shall be replaced with a direct re-rank instruction. We also add additional formatting instructions in the template for post-processing purposes. The detailed, complete prompt template is summarized in \huimin{Appendix~\ref{app:1}}.

\paragraph{Post-processing.} As mentioned, we do not access the model weights or output logits of LLMs. The re-rank results are generated in free texts (e.g., item titles) despite the formatting instructions in the prompts. Therefore, we apply fuzzy matching to transform the generated textual recommendations (e.g., item titles) into a ranked list of item IDs: $\hat{\mathcal{I}}_{RAG}$ to perform evaluation. 
\section{Experiments}
\label{sec:experiments}

\subsection{Experimental Setup}
\begin{table*}[!bht]
\centering
\resizebox{\linewidth}{!}{
\begin{tabular}{@{}llcccccccccccc@{}}
\toprule
\multirow{2}{*}{\textbf{Dataset}} & \multirow{2}{*}{\textbf{Metric}} & \multicolumn{3}{c}{\textbf{ID-Based}} & \multicolumn{4}{c}{\textbf{Text-Based}} & \multicolumn{2}{c}{\textbf{Hybrid}} \\ 
\cmidrule(lr){3-5} \cmidrule(lr){6-9} \cmidrule(l){10-11}
&                                  & FedSAS     & FedLRU     & CF-FedSR     & TransFR    & P5           & RecFmr.                 & UniSR$_{\mathrm{T}}$    & UniSR$_{\mathrm{IT}}$ & \ours     \\ 
\midrule
\multirow{4}{*}{\hfill {\textbf{Beauty}}}
& $\mathrm{R@}5      \uparrow$     & 0.0153     & 0.0218                    & 0.0204                & 0.0059     & 0.0013     & \underline{0.0313}      & 0.0231    & 0.0247                & \textbf{0.0348}                  \\
& $\mathrm{N@}5      \uparrow$     & 0.0101     & 0.0145                    & 0.0138                & 0.0037     & 0.0007     & \underline{0.0167}      & 0.0159    & 0.0166                & \textbf{0.0233}                  \\
& $\mathrm{R@}10     \uparrow$     & 0.0241     & 0.0336                    & 0.0306                & 0.0088     & 0.0024     & \underline{0.0533}      & 0.0347    & 0.0349                & \textbf{0.0563}                  \\
& $\mathrm{N@}10     \uparrow$     & 0.0129     & 0.0183                    & 0.0170                & 0.0047     & 0.0011     & \underline{0.0238}      & 0.0196    & 0.0199                & \textbf{0.0302}                  \\
\midrule
\multirow{4}{*}{{\hfill \textbf{Games}}}
& $\mathrm{R@}5      \uparrow$     & 0.0289     & 0.0391                    & 0.0366                & 0.0059     & 0.0019     & 0.0399                  & 0.0327    & \underline{0.0412}    & \textbf{0.0471}                  \\
& $\mathrm{N@}5      \uparrow$     & 0.0195     & 0.0257                    & 0.0235                & 0.0037     & 0.0010     & 0.0227                  & 0.0218    & \underline{0.0274}    & \textbf{0.0331}                  \\
& $\mathrm{R@}10     \uparrow$     & 0.0419     & 0.0613                    & 0.0560                & 0.0096     & 0.0042     & \underline{0.0706}      & 0.0522    & 0.0621                & \textbf{0.0764}                  \\
& $\mathrm{N@}10     \uparrow$     & 0.0237     & 0.0329                    & 0.0298                & 0.0049     & 0.0017     & 0.0326                  & 0.0281    & \underline{0.0342}    & \textbf{0.0406}                  \\
\midrule
 \multirow{4}{*}{{\hfill \textbf{Toys}}}
& $\mathrm{R@}5      \uparrow$     & 0.0094     & 0.0182                    & 0.0162                & 0.0067     & 0.0156     & \textbf{0.0476}         & 0.0323    & 0.0183                & \underline{0.0419}               \\
& $\mathrm{N@}5      \uparrow$     & 0.0070     & 0.0133                    & 0.0125                & 0.0043     & 0.0080     & \underline{0.0250}      & 0.0225    & 0.0125                & \textbf{0.0268}                  \\
& $\mathrm{R@}10     \uparrow$     & 0.0124     & 0.0243                    & 0.0211                & 0.0106     & 0.0240     & \textbf{0.0739}         & 0.0495    & 0.0271                & \underline{0.0720}               \\
& $\mathrm{N@}10     \uparrow$     & 0.0080     & 0.0153                    & 0.0141                & 0.0055     & 0.0107     & \underline{0.0355}      & 0.0280    & 0.0155                & \textbf{0.0364}                  \\
\midrule
 \multirow{4}{*}{{\hfill \textbf{Auto}}}
& $\mathrm{R@}5      \uparrow$     & 0.0214     & \underline{0.0607}        & 0.0464                & 0.0107     & 0.0060     & 0.0536                  & 0.0500     & 0.0464               & \textbf{0.0643}                  \\
& $\mathrm{N@}5      \uparrow$     & 0.0138     & 0.0341                    & 0.0294                & 0.0072     & 0.0027     & 0.0264                  & 0.0324     & \underline{0.0372}   & \textbf{0.0390}                  \\
& $\mathrm{R@}10     \uparrow$     & 0.0357     & 0.0786                    & 0.0679                & 0.0143     & 0.0083     & 0.0750                  & 0.0679     & \underline{0.0821}   & \textbf{0.0964}                  \\
& $\mathrm{N@}10     \uparrow$     & 0.0187     & 0.0398                    & 0.0361                & 0.0085     & 0.0035     & 0.0332                  & 0.0384     & \underline{0.0488}   & \textbf{0.0492}                  \\
\midrule
\multirow{4}{*}{{\hfill \textbf{ML-100K}}}
& $\mathrm{R@}5      \uparrow$     & 0.0183     & 0.0459                    & \underline{0.0550}    & 0.0092     & 0.0002     & 0.0001                  & 0.0183     & 0.0183               & \textbf{0.0642}                  \\
& $\mathrm{N@}5      \uparrow$     & 0.0085     & 0.0327                    & \textbf{0.0402}    & 0.0035     & 0.0001     & 0.0001                  & 0.0150     & 0.0075               & \underline{0.0362}                  \\
& $\mathrm{R@}10     \uparrow$     & 0.0367     & \underline{0.1101}        & 0.1009                & 0.0092     & 0.0002     & 0.0091                  & 0.0183     & 0.0459               & \textbf{0.1468}                  \\
& $\mathrm{N@}10     \uparrow$     & 0.0145     & 0.0538                    & \underline{0.0550}    & 0.0035     & 0.0001     & 0.0031                  & 0.0150     & 0.0166               & \textbf{0.0621}                  \\
\midrule
\multirow{4}{*}{{\hfill \textbf{Average}}}
& $\mathrm{R@}5      \uparrow$     & 0.0187     & \underline{0.0371}        & 0.0350                & 0.0077     & 0.0050     & 0.0345                  & 0.0313     & 0.0298               & \textbf{0.0505}                  \\
& $\mathrm{N@}5      \uparrow$     & 0.0118     & \underline{0.0241}        & 0.0239                & 0.0045     & 0.0025     & 0.0182                  & 0.0215     & 0.0202               & \textbf{0.0313}                  \\
& $\mathrm{R@}10     \uparrow$     & 0.0302     & \underline{0.0616}        & 0.0553                & 0.0105     & 0.0070     & 0.0546                  & 0.0445     & 0.0504               & \textbf{0.0896}                  \\
& $\mathrm{N@}10     \uparrow$     & 0.0156     & \underline{0.0320}        & 0.0304                & 0.0054     & 0.0034     & 0.0256                  & 0.0258     & 0.0270               & \textbf{0.0437}                  \\
\bottomrule
\end{tabular}
}
\caption{Main results on recommendation performance under different FR schemes. The best results are highlighted in bold and the second best results are highlighted with underline.}
\label{tab:main}
\end{table*}

\paragraph{Datasets.} We select 5 benchmark datasets from different domains to evaluate \ours. They are Beauty, Games, Toys, Auto \cite{he2016ups,mcauley2015image} and ML-100K \cite{harper2015movielens}. The first four datasets are Amazon review datasets consisting of user feedback on different categories of products. ML-100K is a benchmark movie recommendation dataset. Following \cite{chen2023palr,yue2022defending}, we pre-process the raw data with 5-core, and construct training sequences in chronological order. The detailed dataset statistics are in \huimin{Table~\ref{tab:overall_stats}} (Appendix~\ref{app:0}). 


\paragraph{Baselines.} We adopt three groups of baseline methods for comparison. (1) Group 1: ID-based FR models, namely FedSAS, FedLRU and CF-FedSR. FedSAS and FedLRU are federated version of SASRec \cite{kang2018self} and LRURec \cite{yue2023linear} with FedAvg \cite{mcmahan2017communication} as the aggregation protocol. CF-FedSR \cite{luo2022towards} uses a client utility-aware aggregation protocol for better FR. (2) Group 2: Text-based models, i.e., TransFR \cite{zhang2024transfr}, P5 \cite{geng2022recommendation} and RecFmr \cite{li2023text} and UniSR$_{\mathrm{T}}$ \cite{hou2022towards}. TransFR trains a BERT-like retriever. UniSR$_{\mathrm{T}}$ trains SASRec using BERT-encoded item texts. P5 and RecFmr finetune pretrained T5 or LongFormer with item texts and use the model as recommender. (3) Group 3: The hybrid UniSR$_{IT}$ \cite{hou2022towards}, an ID-augmented version of UniSR$_{\mathrm{T}}$. Note that P5, RecFmr, UniSR$_{T}$ and UniSR$_{IT}$ are not originally designed for FR. We use FedAvg as the aggregation protocol to extend them into our FR setting. We did not include LLM-based recommenders as baselines, because existing LLM-based recommenders solely focus on the ranking stage, and could not provide a complete solution for FR: they can only rank ground-truth items along with other sampled negative items. The can not generate candidates from the entire item scope from scratch, which is required by FR.

\paragraph{FR Setup.} Since the training of some baseline methods could not converge under a large number of clients, we compare all methods under a setting of 5 clients. Moreover, to simulate data sparsity and data heterogeneity, the local datasets are sparsely sampled from the original datasets and are guaranteed to be mutually exclusive in terms of users. The remaining users are used as cold-start test users. The detailed local datasets statistics are summarized in \huimin{Table~\ref{tab:fl_stats}} (Appendix~\ref{app:0}). Note that the test users in our FR setting are cold-start users, \huimin{whose historical data are not used for model training}. This is fundamentally different from the common centralized sequential recommendation setting, where the historical data of all users are used to train the model with the last item used for testing. Our cold-start setting is more realistic in the FR setting, and also more appropriate to test the generalization of the trained FR model.

\paragraph{Implementation.} To implement \ours, we train LRURec from scratch, and finetune the pretrained E5 (\texttt{e5-base-v2}) using the training data. When training LRURec, the learning rate is initialized as 1e-3, and the number of local/global epochs is 80/5 (for Beauty, Games, Toys) and 60/5 (for Auto and ML-100k). As for E5, the learning rate is initialized as 1e-6, and the number of local/global epochs is 2/2. We use AdamW as the optimizer for both LRURec and E5. When generating the candidate set $\hat{\mathcal{I}}$ in the first stage, we pick the top-20 items from the hybrid score $\hat{\mathcal{P}}_{hybrid}$. To evaluate the recommendation performance, we select the commonly used normalized discounted cumulative gain (NDCG@N) and recall (Recall@N)
with $N \in [5, 10]$. The predictions are ranked against all items in the dataset. More implementation details on baseline methods are in \huimin{Appendix~\ref{app:0}}.

\subsection{Evaluation: Overall Recommendation Performance}
The first set of results are reported in Table~\ref{tab:main}, where we compare the recommendation performance of all schemes. In Table~\ref{tab:main}, for clarity, we highlight the best results in bold and underline the second best results. Note that the ranking results are from the entire item scope (i.e., a complete ranking of all items).  From Table~\ref{tab:main}, it is observed: (1) \ours generally achieves better recommendation performance than the baseline methods. For instance, compared to the second best baseline method over all datasets, \huimin{\ours achieves 36.12\%, 29.88\%, 45.44\% and 36.56\% average improvements w.r.t. all 4 metrics, respectively.} (2) \ours achieves satisfying performance on ML-100K, \huimin{whereas some baseline methods, especially text-based ones, could barely converge.} This is expected, because some of them (e.g., RecFmr, UniSR) are pretrained using Amazon review datasets, whose domain is different from the domain of movie recommendation. Moreover, the user history of ML-100K is much longer than that in Amazon review datasets. Since the \huimin{text-based} baseline methods could not handle long user history, they are prone to suffer from performance degradation. On the contrary, \ours leverages hybrid retrieval and exploits the pre-trained knowledge in LLMs, thus enjoying a more comprehensive understanding of user preferences. 

\subsection{Sensitivity Analysis}
\begin{figure}[t]
\centering
\begin{subfigure}{\textwidth}
\centering
\includegraphics[trim=0cm 0cm 0cm 0cm, clip, width=0.8\textwidth]{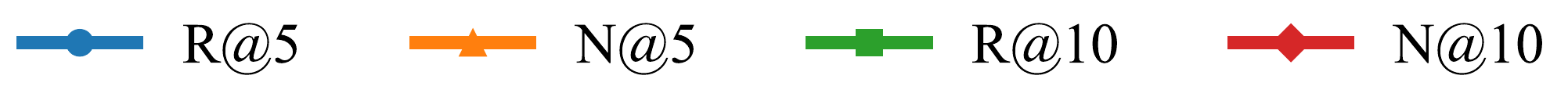}
\end{subfigure}
\begin{subfigure}{0.49\textwidth}
\includegraphics[trim=0.0cm 0.0cm 0.0cm 2.2cm, clip, width=\columnwidth]{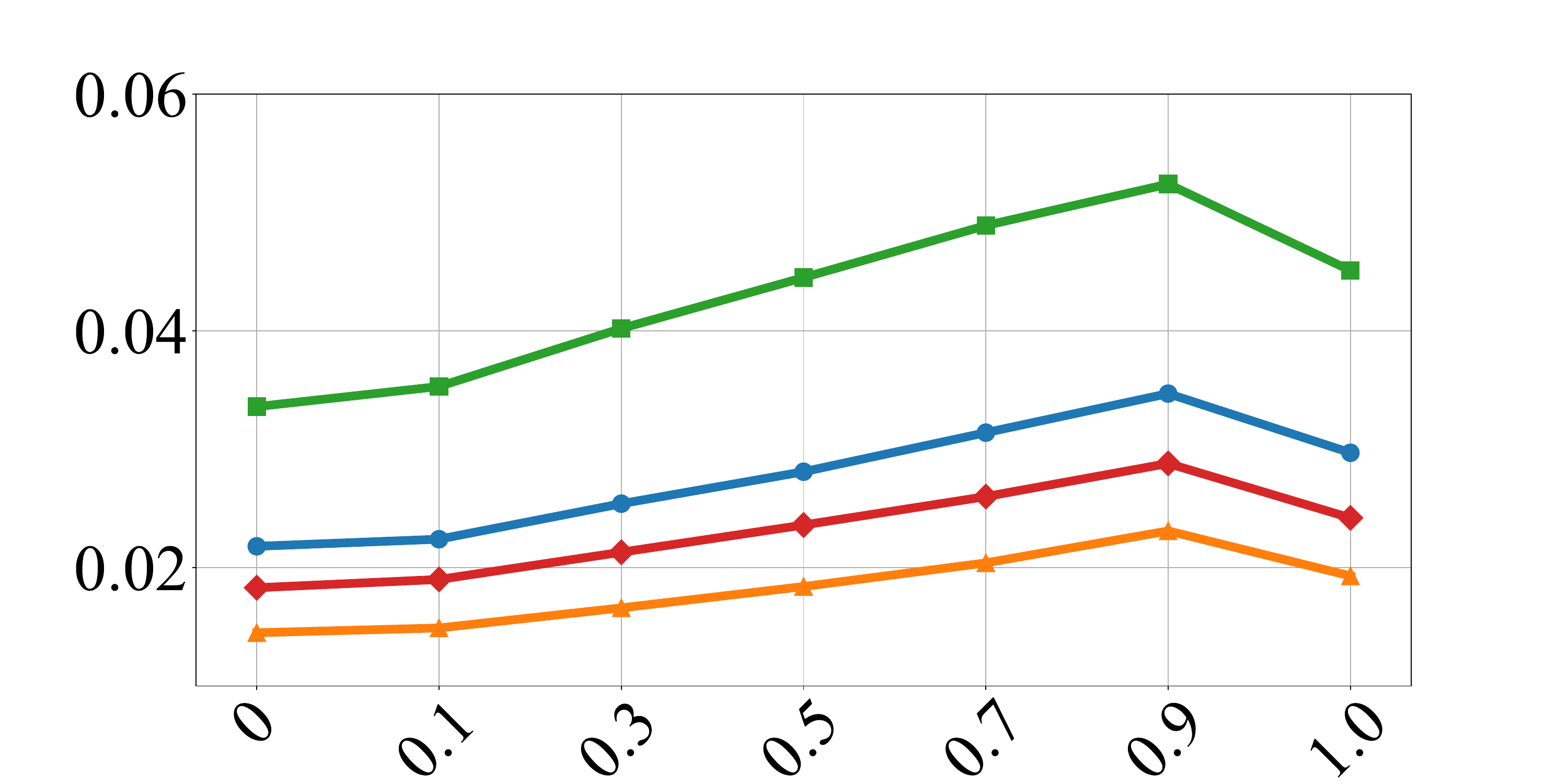}
\caption{Beauty}
\end{subfigure}
\begin{subfigure}{0.49\textwidth}
\includegraphics[trim=0.0cm 0.0cm 0.4cm 2.2cm, clip, width=\columnwidth]{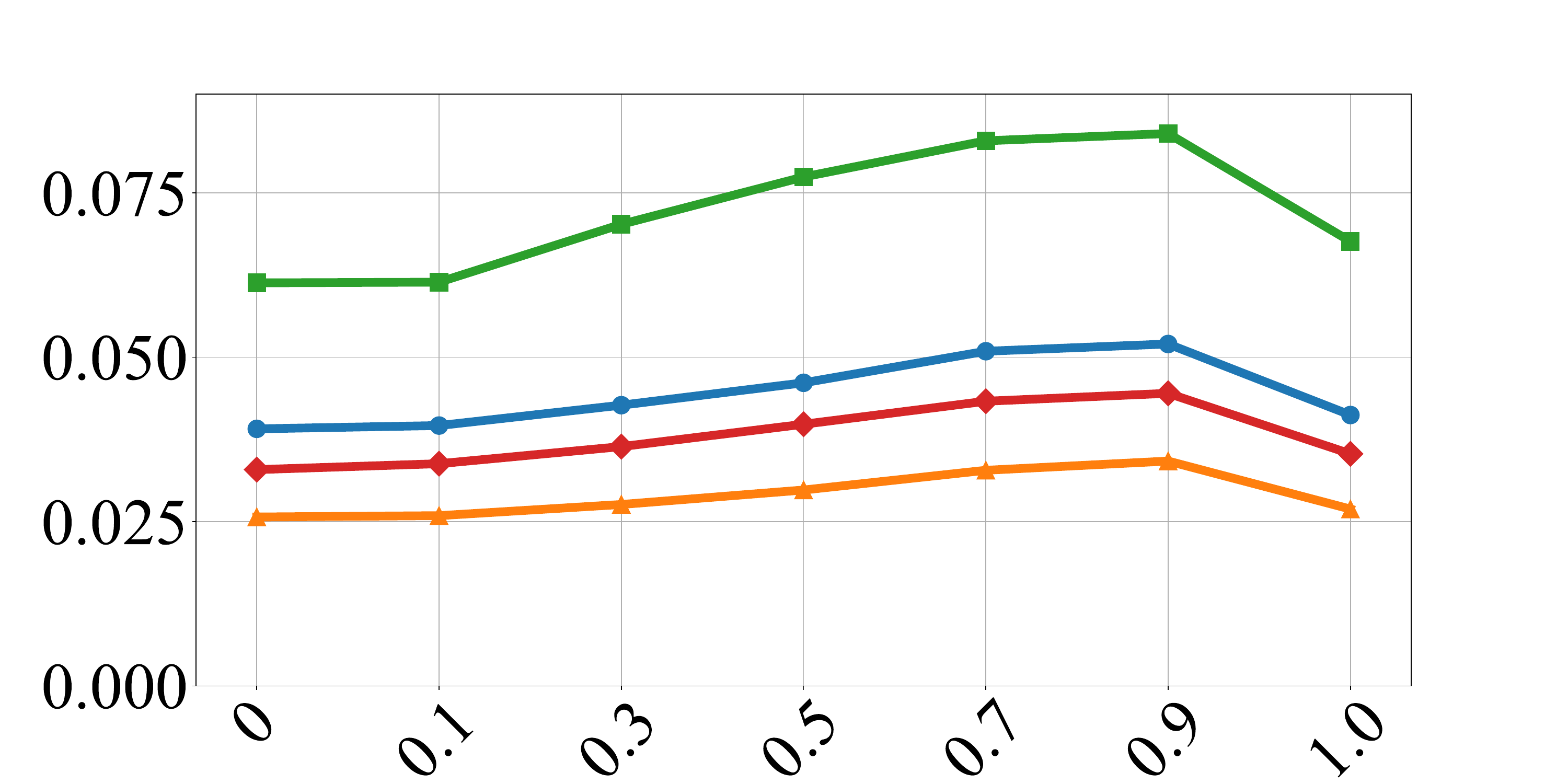}
\caption{Games}
\end{subfigure}
\begin{subfigure}{0.49\textwidth}
\includegraphics[trim=0.0cm 0.0cm 0.0cm 2.2cm, clip, width=\columnwidth]{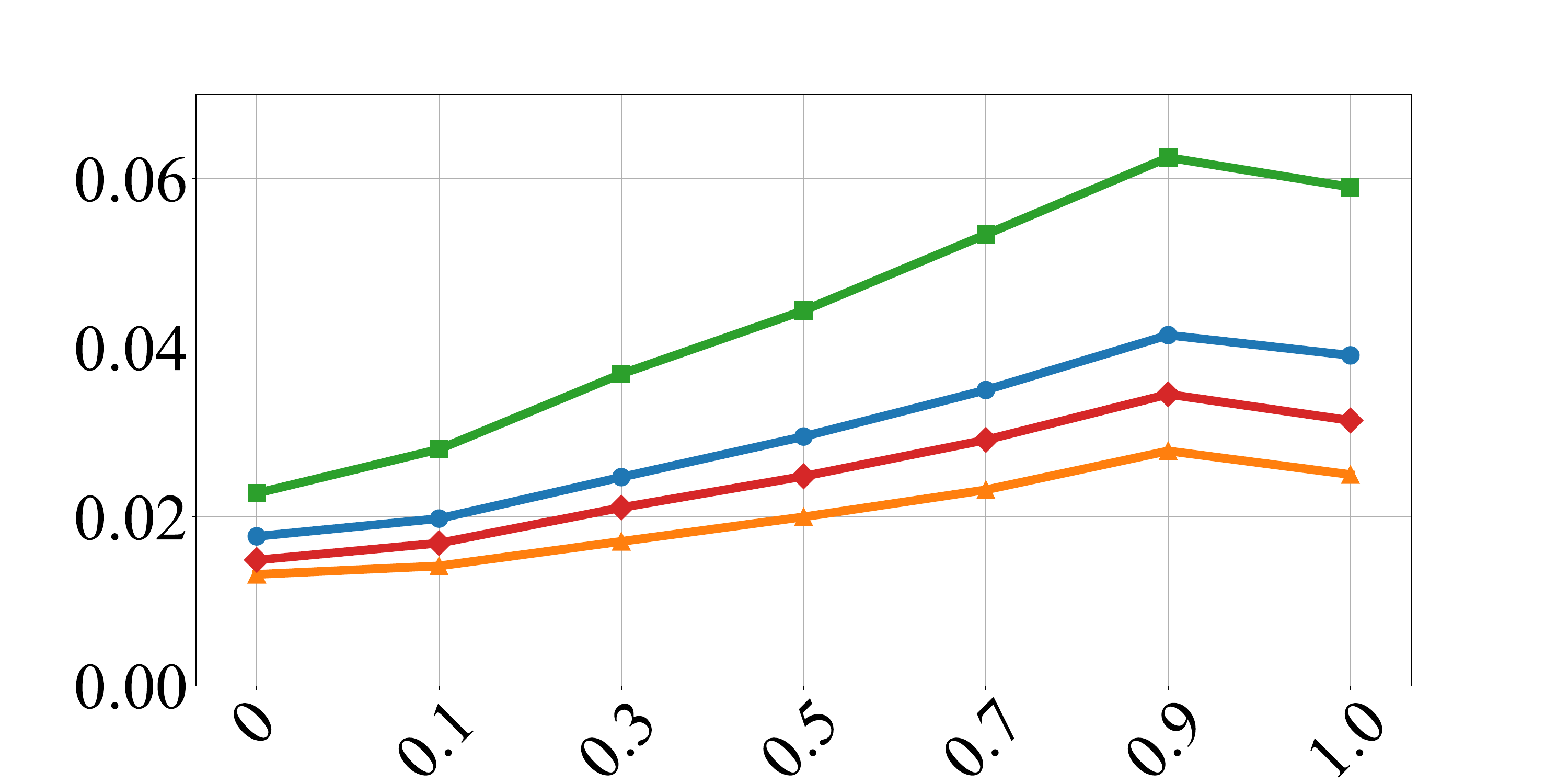}
\caption{Toys}
\end{subfigure}
\begin{subfigure}{0.49\textwidth}
\includegraphics[trim=0.0cm 0.0cm 0.4cm 2.2cm, clip, width=\columnwidth]{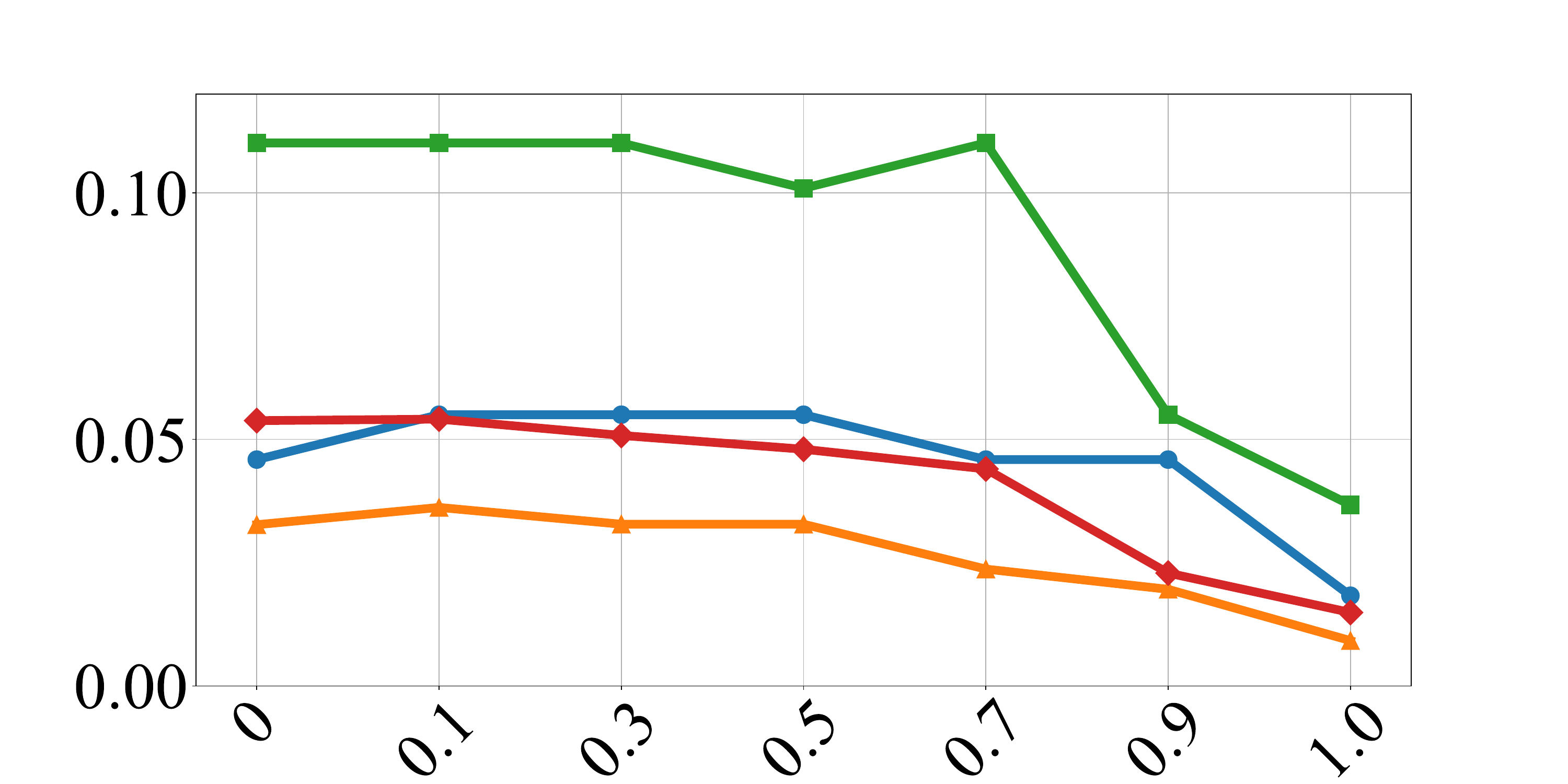}
\caption{ML-100K}
\end{subfigure}
\caption{Sensitivity Analysis. We change $\lambda$ in Equation~\ref{eq:hybrid}, and evaluate the recommendation performance based on the hybrid retrieval results of stage 1.}
\label{fig:sensitivity}
\end{figure}

In this section, we conduct a sensitivity analysis w.r.t. $\alpha$ in Equation~\ref{eq:hybrid}\footnote{Due to the limit budget of calling GPT API, we only perform sensitivity analysis for the first stage of \ours.}. This is a key factor affecting the hybrid retrieval before LLM. In particular, we change it from $0$ to $1$ and keep other configurations the same. The results are shown in Figure~\ref{fig:sensitivity}. We observe that the balance between ID-based retrieval and text-based retrieval play an important role in terms of hybrid retrieval. \huimin{As shown in Figure~\ref{fig:sensitivity}, there exists an optimal $\alpha$ between $0$ and $1$ for different datasets respectively, indicating the necessity of hybrid retrieval in \ours.} 

\subsection{Ablation Study}

\begin{table}[t]
\centering
\resizebox{1.05\textwidth}{!}{
\begin{tabular}{lcccccccccc}
\toprule
\textbf{Dataset} & \textbf{Metric}    & \textbf{\ours}  & w/o LRU   & w/o E5    & w/o LLM   \\ 
\midrule
\multirow{4}{*}{\textbf{Beauty}}
& $\mathrm{R@}5      \uparrow$     & 0.0348     & 0.0306    & 0.0254  & 0.0348 \\
& $\mathrm{N@}5      \uparrow$     & 0.0233     & 0.0203    & 0.0156  & 0.0232 \\
& $\mathrm{R@}10     \uparrow$     & 0.0563     & 0.0504    & 0.0478  & 0.0520 \\
& $\mathrm{N@}10     \uparrow$     & 0.0302     & 0.0267    & 0.0227  & 0.0288 \\
\midrule
\multirow{4}{*}{\textbf{Games}}
& $\mathrm{R@}5      \uparrow$     & 0.0471     & 0.0425    & 0.0438  & 0.0435 \\
& $\mathrm{N@}5      \uparrow$     & 0.0331     & 0.0259    & 0.0290  & 0.0281 \\
& $\mathrm{R@}10     \uparrow$     & 0.0764     & 0.0779    & 0.0722  & 0.0689 \\
& $\mathrm{N@}10     \uparrow$     & 0.0406     & 0.0372    & 0.0381  & 0.0364 \\
\bottomrule
\end{tabular}
}
\caption{Ablation Study. We mask out different components of \ours separately, and evaluate the recommendation performance.}
\label{tab:ablation}
\end{table}

Finally, we conduct an ablation study to evaluate the contribution of each key module of \ours, namely the hybrid retrieval in the first stage and the LLM-based re-ranking in the second stage. The results are reported in Table~\ref{tab:ablation}. In Table~\ref{tab:ablation}, w/o LRU refers to \ours only with E5 as retriever in the first stage, and w/o E5 indicates that there is only LRURec in the first stage. w/o LLM represents \ours without the LLM-based re-rank. As expected, we observe that both hybrid retrieval and LLM-based re-ranking contributes to a better recommendation performance. For instance, without either module, the recommendation performance basically drops. Such results also validate the merit of our design: our hybrid retrieval mechanism and the LLM-based re-ranking are indeed effective in terms of addressing the data sparsity and data heterogeneity in FR, which leads to better recommendation performance.

\section{Conclusion}
\label{sec:conclu}
This work presents a novel federated recommendation framework that exploits ChatGPT and a novel hybrid retrieval mechanism. \ours provides an effective privacy-aware solution to build a recommender system for data-sparse and data-heterogeneous federated recommendation scenarios. We highlight the significance of this work: despite the active research on federated recommendation, existing methods largely suffer from the data sparsity and data heterogeneity issue in FR. In contrast, our work is deliberately designed to overcome such an issue, achieving generalized recommendation performance. Finally, within \ours, there is also a hybrid RAG mechanism to prevent LLM hallucination, improving the authenticity of recommendation results in real-world applications.

\section{Limitations}
\label{sec:limitations}
One limitation of this work is that our method introduces extra hyperparameters. For different applications, one might need to finetune these hyperparameters, which brings extra computational cost. Another limitation of this work is that our method does not take the inherent bias of GPT into account. However, it is known that such pretrained LLMs usually have encoded the bias in the pre-training data (e.g., stereotypical data, racism and hate speech). Such bias could have negative ethical implications on the downstream FR applications. Therefore, a future research direction is to develop a benign and fair FR framework. 



\bibliography{reference}

\onecolumn
\appendix
\section{Implementation Details}
\label{app:0}
\subsection{Datasets Details and FR Setup}
\paragraph{Overall Dataset Statistics.} The detailed dataset statistics (after 5-core processing) is reported in Table~\ref{tab:overall_stats}. According to Table~\ref{tab:overall_stats}, it is observed that the averaged sequence length in ML-100K is significantly larger than that of other datasets. In addition, the density of ML-100K is also much higher than others. Finally, the Beauty, Games, Toys and Auto datasets are all Amazon review datasets, whose domain is rather different from that of ML-100K.

\begin{table}[h]
\centering
\resizebox{0.55\textwidth}{!}{
\begin{tabular}{lcccccccccc}
\toprule
\textbf{Datasets}     & \textbf{users}      & \textbf{Items}    &\textbf{Interaction}   &\textbf{Length}     &\textbf{Density}  \\ 
\midrule
\textbf{Beauty}       & 22332               & 12086             & 198K                  & 8.88               & 7e-4             \\
\textbf{Games}        & 15264               & 7676              & 147K                  & 9.69               & 1e-3             \\
\textbf{Toys}         & 19412               & 11924             & 167K                  & 8.63               & 7e-3             \\
\textbf{Auto}         & 1281                & 844               & 8K                    & 6.70               & 8e-3             \\
\textbf{ML-100K}      & 610                 & 3650              & 89K                   & 146.70             & 4e-2             \\
\bottomrule
\end{tabular}
}
\caption{Overall dataset statistics.}
\label{tab:overall_stats}
\end{table}

\paragraph{Federated Dataset Statistics.} In our experiments, we randomly sample a set of users from the original datasets, and randomly distributed them onto different local clients. In terms of users, it is guaranteed that the local datasets are mutually exclusive: one user could one exist in one local dataset or only in the test dataset. The per-client statistics is summarized in Table~\ref{tab:fl_stats}. Note that considering the dataset size and training convergence, we sampled different number of training users for different datasets. For each dataset, after sampling the training users, the remaining users are used for validation and evaluation. It is observed that for Beauty, Games and Toys, the test item scope is significantly larger than each local item scope, leading to a data sparse and data heterogeneous FR setting.

\begin{table}[h]
\centering
\resizebox{0.6\textwidth}{!}{
\begin{tabular}{lcccccccccc}
\toprule
\textbf{Datasets}     & \textbf{Users/Client}   & \textbf{Items/Client}   &\textbf{Test Users}  &\textbf{Test Items}   \\ 
\midrule    
\textbf{Beauty}       & 1000                    & 4161                    & 17332               & 12086                \\
\textbf{Games}        & 1000                    & 3826                    & 10264               & 7614                 \\
\textbf{Toys}         & 1000                    & 4412                    & 14412               & 11766                \\
\textbf{Auto}         & 200                     & 503                     & 281                 & 591                  \\
\textbf{ML-100K}      & 100                     & 3134                    & 109                 & 3060                 \\
\bottomrule
\end{tabular}
}
\caption{Averaged statistics on local clients and the statistics of test users.}
\label{tab:fl_stats}
\end{table}

\subsection{Implementation.} 
\paragraph{\ours.} To implement \ours, we train LRURec from scratch, and finetune the pretrained E5 (\texttt{e5-base-v2}) using the training data. When training LRURec, the learning rate is initialized as 1e-3, and the number of local/global epochs is 80/5 (for Beauty, Games, Toys) and 60/5 (for Auto and ML-100k). As for E5, the learning rate is initialized as 1e-6, and the number of local/global epochs is 2/2. We use AdamW as the optimizer for both LRURec and E5. When generating the candidate set $\hat{\mathcal{I}}$ in the first stage, we pick the top-20 items from the hybrid score $\hat{\mathcal{P}}$. 

In the second stage, when performing the LLM-based re-ranking, the ideal procedure is to re-rank the candidate sets of all users. However, due to the limited GPT API query budget, such an ideal evaluation procedure is too expensive and infeasible. Moreover, since we are only interested in the recommendation performance w.r.t. the top-20 items, it is equivalent and sufficient to only perform the LLM re-ranking for a subset of test users, whose candidate set includes the ground-truth item. This procedure is meaningful and fair. To see this, we emphasize that we only feed the predicted top-20 items into LLM for re-ranking. In this setting, for a test user, if the ground-truth item is not within these predicted top-20 items, then the ground-truth item will not participate in the LLM re-ranking. As such, re-ranking will not affect the ranking of the ground-truth item, and therefore, will not change the values of Recall@5, Recall@10, NDCG@5 and NDCG@10 either. Technically, to determine whether the recommendation results of a test user should be re-ranked or not, we first validate whether the ground-truth item is included its predicted candidate set (i.e., top-20 items). That is, for a test user, if the ground-truth item is within the predicted top-20 items, then the predicted top-20 items will be fed into LLM for re-ranking. 

Finally, when post-processing the generated texts, we also filter out the re-ranked results without ground-truth items. This procedure makes sense, because the re-ranking process is expected to explicitly focuses on re-ranking the top-20 items with ground-truth items in them. However, if a generated re-ranked item list does not contain the ground-truth item, we ignore this re-ranked list and use the candidate set from stage one as the final recommendation for this test user. For instance, we observe that GPT-3.5 may ignore some technical parameters within some product titles or abbreviate some product titles. This leads to a discrepancy between the true titles of the ground-truth items and the generated ones, even if in human eyes they may represent the same product. Such generated results are treated as noisy generation and are ignored when calculating the evaluation metrics in our evaluation.

\paragraph{Baselines.} For all baseline methods, we refer to the original papers and the official implementations. Moreover, for non-FR methods, we used the same code of FedAvg used for \ours to perform global model aggregation. For all baselines, we train the models by starting with the default training/finetuning configuration. For FedSAS, FedLRU, CF-FedSR and TransFR, the models are trained from scratch. In comparison, we load the released pretrained weights for P5, RecFmr and UniSR, and finetune them in our FR setting. Finally, we observed training divergence and overfitting of some baseline methods, and therefore, adjusted both local epochs and global epoch until find the best performance.

\clearpage

\section{Prompt Engineering}
\label{app:1}
\paragraph{E5 Templates.} We follow the instructions in \cite{wang2022text}, and design our E5 templates as follows:
\begin{displayquote}
\#\#\# query: \{\texttt{history}\}

\#\#\# passage: \{\texttt{candidate}\},
\end{displayquote}
where \texttt{history} shall be replaced with history item titles and genres/categories. \texttt{candidate} shall be replaced with the title and category/genre of a single candidate item. Moreover, in our experiments, we notice that it is more beneficial to only use last several movies for the ML-100K dataset, and does not use genre. For instance, an exemplar complete template on ML-100K is like (note that the history does not need to be the complete history and the description of genres may be removed for better performance):

\begin{displayquote}
\#\#\# query: 

\texttt{The Shawshank Redemption, a movie about Thriller};

\texttt{Ex Machina , a movie about Sci-Fi, Thriller};

\texttt{Unchained, a movie about Drama, Western}.

\#\#\# passage: 

\texttt{Whiplash, a movie about Drama}.
\end{displayquote}

\paragraph{GPT-3.5-Turbo Templates.} Inspired by \cite{hou2023large}, we use the following templates to prompt GPT-3.5-Turbo:

\begin{displayquote}
\#\#\# system: 

You are a helpful \{\texttt{role}\},  \{\texttt{role description}\}.

\#\#\# user: 

I've browsed the following items in the past in order: 

\{\texttt{history}\}. 

There is also candidate pool: 

\{\texttt{candidates}\}. 

\{\texttt{task instruction}\}. 
\end{displayquote}

In the above template, \texttt{role} and \texttt{role description} are replaced with a concrete real-world job (e.g., movie reviewer) and its job description (e.g., recommending movies for people). \texttt{history} is instantiated as item texts containing item titles and item categories/genres. As for \texttt{candidates}, it is replaced with the hybrid retrieval results. \texttt{task instruction} shall be replaced with a direct re-rank order. We also add additional formatting instructions in the template for post-processing purposes. An exemplar prompt for ML-100K is like:

\begin{displayquote}
\#\#\# system: 

You are a \texttt{movie fan and movie reviewer}. Therefore, people might ask you to \texttt{recommend movies.}

\#\#\# user: 

I've browsed the following items in the past in order: 

\texttt{The Shawshank Redemption, a movie about Thriller};

\texttt{Ex Machina , a movie about Sci-Fi, Thriller};

\texttt{Unchained, a movie about Drama, Western}.

There is also candidate pool: 

\texttt{1. The Green Mile (1999)}

\texttt{2. Pulp Fiction (1994)}

\texttt{3. Seven (1995)}

\texttt{Please rank these movies by measuring the possibilities that I would like to watch next most, according to my watching history. Please think step by step.
Please show me your ranking results with order numbers. Split your output with line break. You MUST rank the given candidate movies. You can not generate movies that are not in the given candidate list}. 
\end{displayquote}

\end{document}